\documentclass[a4paper]{article}
\usepackage[english]{babel}
\usepackage{icrc2013}
\title{Comparison of Different Trigger and Readout Approaches for Cameras in the Cherenkov Telescope Array Project }

\shorttitle{"Comparison of Different Trigger and Readout Approaches for Cameras in the CTA Project"}

\authors{
M. Shayduk$^{1}$,
S. Vorobiov$^{2}$,
U. Schwanke$^{2}$,
R. Wischnewski$^{1}$
for the CTA Consortium.
}

\afiliations{
$^1$ DESY Zeuthen\\
$^2$ Humboldt University, Berlin \\

}

\email{maxim.shayduk@desy.de}

\abstract{
The Cherenkov Telescope Array (CTA) is a next-generation ground-based observatory for $\gamma$-rays 
with energies between some ten GeV and a few hundred TeV. CTA is currently in the advanced design phase 
and will consist of arrays with different size of prime-focus Cherenkov telescopes, to ensure a proper energy 
coverage from the threshold up to the highest energies. The extension of the CTA array with double-mirror 
Schwarzschild-Couder telescopes is planned to improve the array angular resolution over wider field of view. 
We present an end-to-end Monte-Carlo comparison of trigger concepts for the different imaging cameras 
that will be used on the Cherenkov telescopes. The comparison comprises three alternative trigger schemes 
(analog, majority, flexible pattern analysis) for each camera design. The study also addresses the influence 
of the properties of the readout system (analog bandwidth of the electronics, length of the readout window in time) 
and uses an offline shower reconstruction to investigate the impact on key performances 
such as energy threshold and flux sensitivity.
}


\begin{document}
\maketitle

\section{Introduction}

The emerging success of the very high energy ground-based $\gamma$-ray astronomy was prominently ensured by the Imaging Cherenkov telescopes technique \cite{bib:whipple2},\cite{bib:hegra}. The currently operating Cherenkov telescope 
experiments like H.E.S.S. \cite{bib:hess}, MAGIC \cite{bib:magic} and VERITAS \cite{bib:veritas} have already proved to be very capable instruments to study the very high energy astrophysical phenomena 
both in our galaxy and in extragalactic sources. The next generation ground-based $\gamma$-ray  experiment - Cherenkov Telescope Array (CTA) Observatory \cite{bib:CTA} is currently in the preparatory phase.
It will provide an order of magnitude higher sensitivity and extend the observable $\gamma$-ray energy range up to hundreds of TeV. The CTA will comprise about 60 Cherenkov telescopes with different sizes of the reflector and will be extended 
with an array of double-mirror telescopes based on Schwarzschild-Coude optical design.


The Monte Carlo studies for CTA done with the first massive production of simulations \cite{bib:ctamc} were following a
conventional majority/next-neighbor logic for the single telescope trigger,
requiring for any pixel that some number of its direct neighbors must have signal of a certain amplitude within a given coincidence time. 
Moreover, the readout system concept and signal extraction methods were rather following the present generation H.E.S.S array.

The current hardware developments in CTA are ongoing under considerations of reliability and cost, while keeping the requested performance. 
These developments are aiming to optimize many of telescope components, that resulted in several design options for telescope triggers and readout systems
. 
The proper implementation of all these novel designs in the Monte Carlo simulations provides important information to 
decide what will be finally built.
\section{Trigger simulation}

The detailed Monte-Carlo simulations of the camera trigger and data acquisition system were performed with \textit{trigsim} package. 
The software package details and results of first studies are presented in \cite{bib:trigsim}. 
There are several trigger designs currently considered in CTA, as described in \cite{bib:ctamc}:

\textit{Majority Trigger:}
Each of the analogue pulses coming from photomultipliers in a predefined overlapping camera region is fed into the comparator,  
which produces the digital signal if the initial pulse exceeds the adjustable reference amplitude.  
Then the sum of these digital signals again can be passed to the comparator with a certain threshold to count the number of pixels 
and issue the final camera trigger

\begin{figure}[!t]
  \centering
  \includegraphics[width=0.49\textwidth]{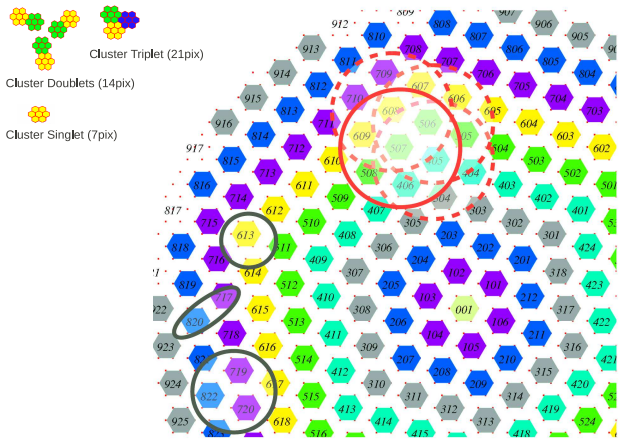}
  \caption{Example of the Middle Size Telescope (MST) camera geometrical structure. Each hexagon represents a 7-pixel-cluster. The dashed red upper
circles indicate 3 neighbored supercluster (each made of 7 next neighbor clusters, corresponding to 49 pixels, which is the region processed by the 
trigger cluster-FPGA.). The baseline trigger patches geometries: Singlets, Doublets and Triplets (contained within grey contours) are shown and also given in the inset.}
  \label{patches_fig}
 \end{figure}

\textit{Analogue Sum Trigger:} 
The analogue pulses from all pixels in a predefined overlapping camera region (see examples in Fig.\ref{patches_fig} ) are added, 
regardless to their amplitude and then compared to the 
reference threshold. In order to avoid that photomultiplier after-pulses  dominate the summed signal, 
the amplitude of analogue pulses is limited to a certain value before summation (clipping threshold).

\textit{Binary Trigger:} 
The analogue signal from the photosensors are passed through comparators at regular time intervals, transforming the camera image to a binary pattern. 
This pattern can be processed with flexible trigger FPGA-based classification algorithms, considering the space and time properties of the data. 
With additional thresholds, the camera image can be converted to the several-bit pattern and the trigger approach can emulate on-line image processing, 
similar to image cleaning procedures used in the  off-line data analysis.

\textit{Digital Trigger:} 
This concept is the interplay between the binary trigger algorithm and the fully analogue approach, described above. The signal from the photosensor is 
digitized by a Flash-ADC and the digital signal is passed to the flexible FPGA-based trigger logic. In this FPGA module, the large variety of trigger logic schemes could be implemented, 
including a digital sum trigger or a digital majority trigger. This trigger approach has an
elegant feature that the trigger Flash-ADC data is used as well as the event data,  reduces the amount of front-end electronics components 
(discriminators and comparators), but currently lacks cost-effective solutions for designs faster than 250MHz. 

All of these concepts were implemented in \textit{trigsim} and studied in detail by comparing telescopes collection areas. 
The influence of the camera electronics analog bandwidth on the $\gamma$-ray collection areas and the optimal analog signal pulse width values were 
discussed in \cite{bib:trigsim}. 
The lowest energy thresholds were obtained for the trigger implementations with the faster pulses, especially for LST part of the array. 
Fort the MST and SST the solutions with slower pulses had tendency to possess $\sim15\%$ gain in trigger efficiency for energies above $\sim$500 GeV. 

The possible basic shapes of camera regions, associated with the trigger patches are presented in Fig.\ref{patches_fig}. In our notation, the one hardware unit is the cluster of 7 pixels. 
The trigger geometrical patches formed by one, two and three such clusters, called "Singlets", "Doublets" and "Triplets" correspondingly. The data from these patches, overlapping as shown in Fig.\ref{patches_fig}, 
is examined by the trigger logic to issue the patch trigger. The final camera trigger is the logical "OR" of all patch triggers. The optimal patch area and shape can differ for 
different trigger concepts and the devoted simulations are ongoing.

 \begin{figure}[!t]
  \centering
  \includegraphics[width=0.45\textwidth]{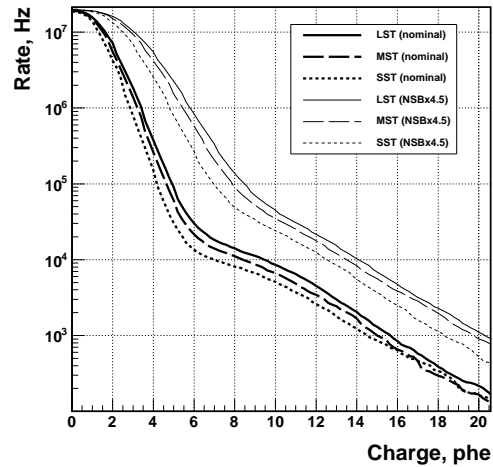}
  \caption{Differential charge spectra in a single pixel for two levels of the night sky background light intensities:  "nominal" denotes the typical intensity level for extra-galactic observations,  
and "NSBx4.5" - galactic/moonlight observation. The "LST", "MST" and "SST" labels stand for Large-, Medium- and Small- Size Telescope accordingly. The single photoelectron pulses are 
gaussian with FWHM=2.6 ns. For high charges (above 6 phe for nominal and 9 phe for high intensity) 
the noise spectrum is fully determined by the photosensor after-pulses. Rate values saturate at $2\cdot10^{7}$Hz due to the 50 ns gate-width of the counter.}
  \label{ipr_fig}
 \end{figure}
 
The trigger threshold of the Cherenkov telescope is determined by the intensity of the night sky background light, which can dramatically vary during observations, depending on the telescope pointing position
and the moon brightness. Moreover, the intrinsic noise of photosensors, so-called after-pulses, contributes to the camera noise rate and affect the telescope performance. Therefore, these aspects should 
be studied as well and the corresponding performances of all trigger concepts should be investigated. 
The Monte-Carlo simulation for the individual pixel noise spectra, induced by the light of the night sky with 
nominal and 4.5 times higher intensity are presented in Fig.\ref{ipr_fig}. 

\begin{figure*}[!t]
 \centering
 \includegraphics[width=0.45\textwidth]{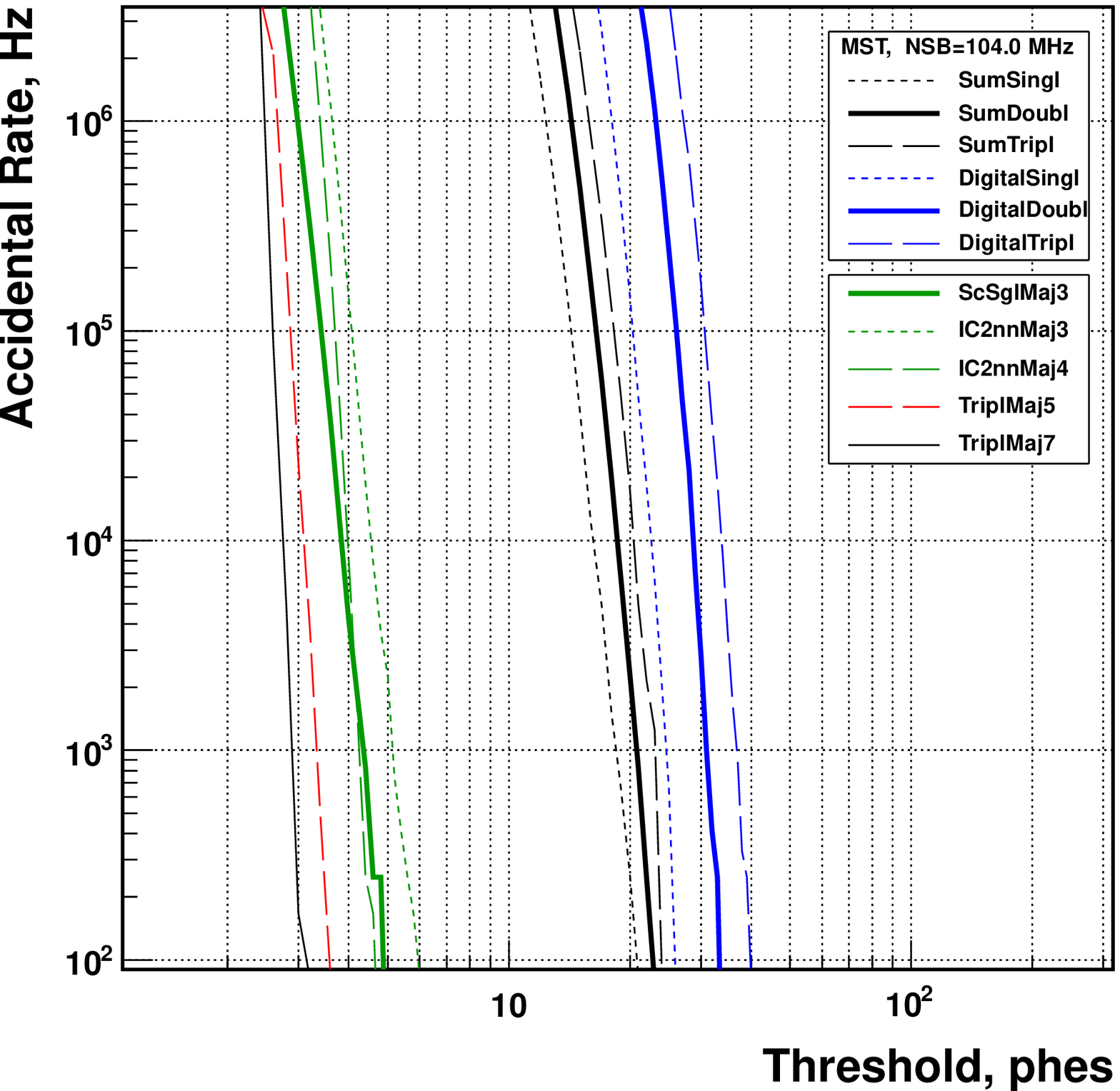}
 \includegraphics[width=0.45\textwidth]{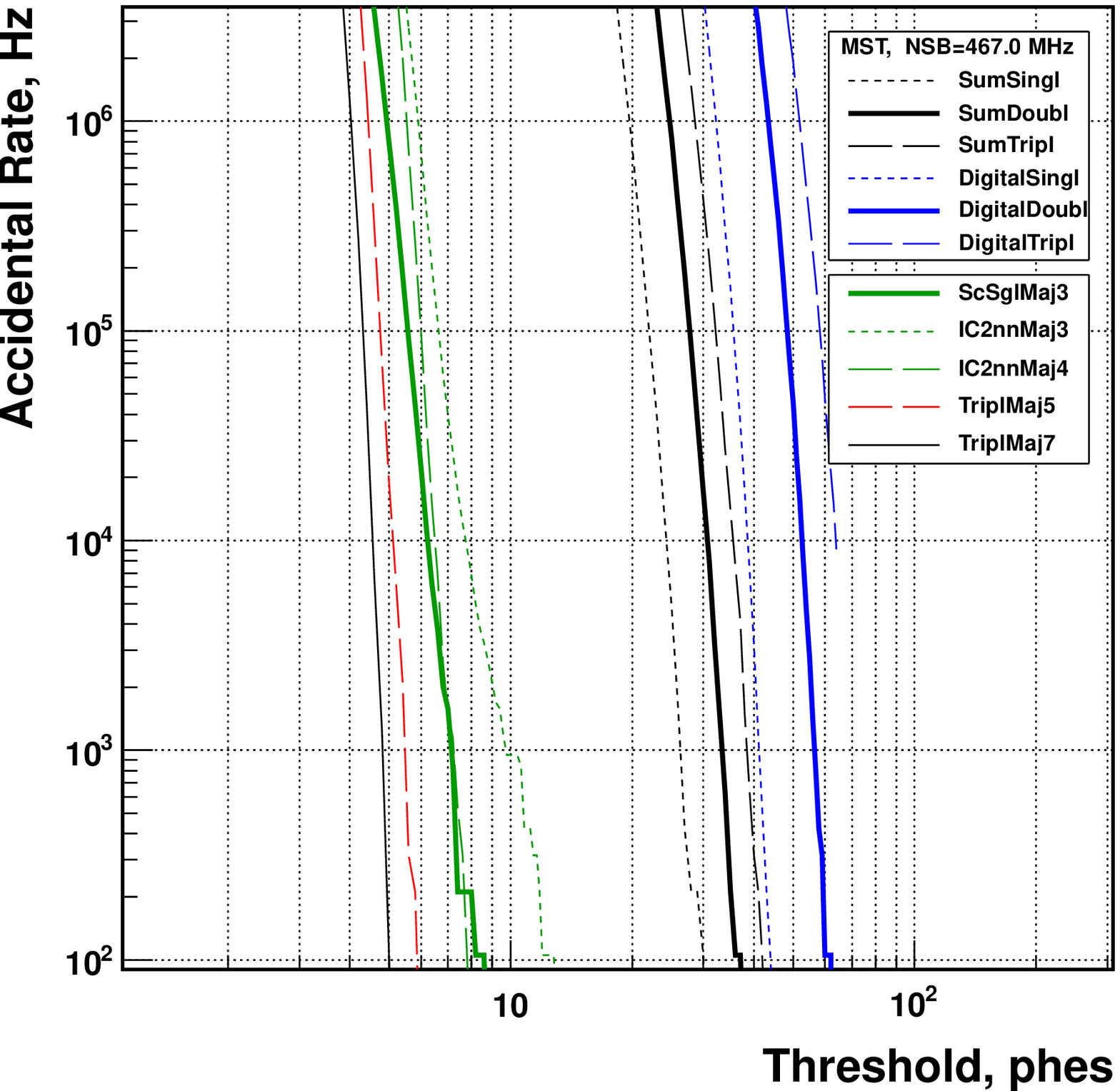}
 \caption{Accidental MST camera rates as a function of  discriminator threshold for different flavors of \textit{Analogue Sum} (labels with prefix "Sum"), \textit{Majority} (ScSgMaj3, IC2nnMaj3, IC2nnMaj4) and \textit{Digital} (labels with prefix "Digital") trigger approaches, 
for standard dark sky NSB rate (left plot) and for 4.5 times larger rate (right plot).  For \textit{Majority} schemes the 
threshold value correspond to the individual pixel discriminator reference amplitude,  while for the \textit{Analogue Sum} and \textit{Digital} approaches the threshold value is for the trigger patch discriminator.
The two-threshold \textit{Binary trigger} approach can be obtained from the majority triggers "IC2nnMaj3" and "IC2nnMaj4" as the logical "OR".}
 \label{CameraRates_fig}
\end{figure*}

The accidental camera trigger rates for all considered trigger concepts with different intrinsic scenarios are shown in Fig.\ref{CameraRates_fig}. 
The camera trigger threshold is defined as the threshold value, corresponding to 10kHz rate.
Following the notation of trigger patches defined in Fig.\ref{patches_fig} the labels "SumSingl", "SumDoubl" and "SumTripl" stand for the \textit{Analogue Sum Trigger} approach with the intrinsic scenario of summing the analogue signals from
the described above trigger Singlets (1 cluster, 7 pixels), Doublets (2 clusters, 14 pixels) and Triplets (3 clusters, 21 pixels).  The single photoelectron  pulse is a gaussian with FWHM=2.6 ns. Similarly, the "DigitalSingl", "DigitalDoubl" and "DigitalTripl" labels denote the digital sum
over the same patches in  \textit{Digital Trigger} concept.  In order to match the 250MHz sampling rate of the Flash-ADC the initial analogue pulse has a special shape, that after FPGA processing is 
roughly equivalent to the gaussian pulse with FWHM$\approx$10 ns (for details, see \cite{bib:flashcam}).   

Trigger schemes listed in lower legends in Fig.\ref{CameraRates_fig} are the fast (analogue pulse FWHM=2.6 ns) \textit{Majority Trigger}  algorithms, designated as following: "ScSglMaj3" - any 3 pixels  out of 7 - pixel cluster should have a signal above a reference threshold within $\sim$1.7 ns coincidence time (each camera pixel can be the center of the cluster),
 "TriplMaj5" and "TriplMaj7" - majority logic with  multiplicities 5 and 7 out of 21-pixel Triplet accordingly. The fancy "IC2nnMaj3"  and "IC2nnMaj4" labels denote the majority trigger algorithms with slightly modified
 logic.  The condition of short coincidence time is  only required for pairs of neighboring pixels, in contrast to the conventional majority scheme, where this condition is examined for all triggered pixels.     
Simultaneous implementation of these two triggers in the FPGA  can serve as the camera two-threshold \textit{Binary Trigger}.

\section{Readout  simulation}

One of the key characteristics of the readout system is the throughput analog bandwidth. Relevantly to the scope of Cherenkov Telescopes, it essentially determines the minimal possible
noise contribution to the recorded signal \cite{bib:vorobiov}. 

We compare here readout system approaches differing basically by the analog bandwidth settings: the high bandwidth possesses  
fast gaussian analogue pulses with 2.6 ns FWHM, digitized  with 1GHz Flash-ADC and the low bandwidth approach with  250MHz Flash-ADC sampling 
rate and slower analogue pulses  with FWHM=10.4 ns. 
For all of these options events were recorded with wide 50 ns default readout window, 
dynamically extended up to 100 ns for events with time duration longer than the default window.
For short integration windows, the reduction of the readout window will allow to lower image cleaning thresholds, since the signal position search range is consiquentely shrinks, 
but potentially can lead to the loss of signal for high-energy events with the large time spread. 
The optimization of readout algorithms in this respect is ongoing \cite{bib:colibri}.   
 
\section{Full analysis chain}

As an input for our simulations we used the data, produced by \textit{simtelarray} program \cite{bib:SimtelArray}.  
This package includes full detector simulations~\cite{bib:ctamc} and provides arrival
times of photoelectrons (p.e.) from showers at the telescope focal planes, relevant for our studies.
Next, in the framework of \textit{trigsim} we simulated the electronic response of the photomultiplier tube (PMT) response by convolving
the p.e. times with the individual single p.e. pulses from the CTA PMT
candidate. The pulses have been widened and sampled according to the bandwidth 
and the sampling rate of the studied front-end electronics designs.

The simulated data have been analyzed using \textit{evndisplay} software package
for the analysis of simulations and data by arrays of imaging Cherenkov telescopes~\cite{bib:EvndispWiki}, originally
developed for VERITAS experiment, and further extended to AGIS and CTA arrays~\cite{bib:maier2007,bib:maier2009}.
We implemented conversion from \textit{sim\_telarray}~\cite{bib:SimtelArray} \textsf{EVENTIO} format to \textit{evndisplay} format
within \textit{trigsim}, in order to analyze all trigger patterns. At the conversion stage, the NSB contribution to Flash-ADC traces is added to all pixels. 

The simulated FADC trace integration, data calibration and the selection of triggered events is being done within \textit{evndisplay}, using trigger thresholds values obtained with \textit{trigsim}. 
For the subsequent image cleaning, the novel image cleaning procedure~\cite{bib:ShaydukIC} has been implemented. 
Then the second-moment parameterization of the recorded images and stereo reconstruction of shower geometry (direction and impact parameter) is performed. 
The $gamma$-ray energy is estimated using beforehand trained lookup tables. 
At the next step we discard events with poor/failed direction reconstruction by applying the dynamic
direction offset cut which corresponds to the typical angular resolution curve for arrays like H.E.S.S. and VERITAS.  

\begin{figure*}[!t]
 \centering
 \includegraphics[width=0.99\textwidth]{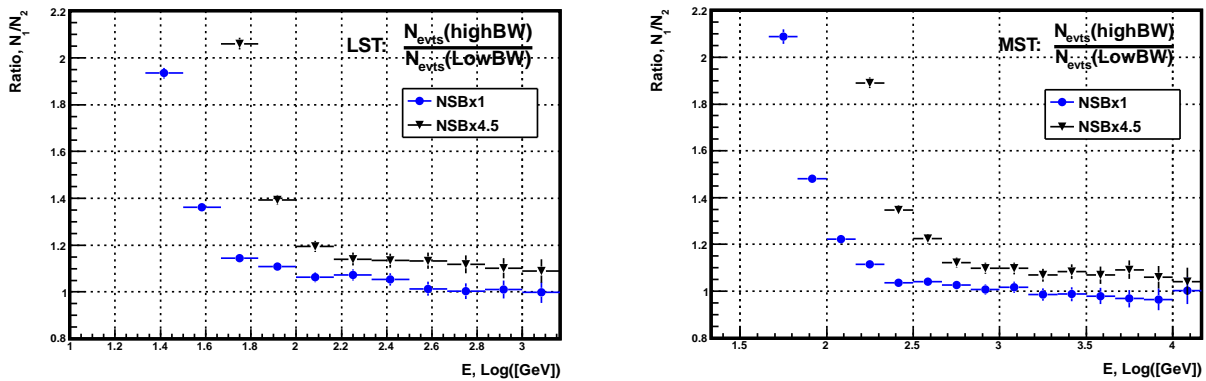}
\caption{Number of reconstructed events after direction cut for the high bandwidth camera approach, normalized to the number of reconstructed events for the  low bandwidth solution. \textbf{Left:}  LST part of array "E". Nominal and high NSB conditions are labelled with circles and triangles correspondingly. \textbf{Right:} MST part of array "E". Significant gain in the number of events in the low energy range for arrays equipped with high BW cameras is revealed. 
}
 \label{SensitivityRatio_fig}
\end{figure*}

\section{Results}

For the comparison of the trigger and readout approaches we selected following camera electronic chains: the high bandwidth chain (denoted in Fig.\ref{SensitivityRatio_fig} as "high BW"), 
equipped with \textit{Analogue Sum} trigger with fast 2.6 ns gaussian pulses and 1GHz Flash-ADC readout system  and the low bandwidth chain ("low BW"), comprising \textit{Analogue Trigger} with slower pulses of 10.4 ns and correspondingly slower 250MHz readout.
It was demonstrated in our previous studies \cite{bib:trigsim}, that the trigger scenario of direct summation of pulses, irrelevant to their amplitudes, leads to lower energy thresholds (for Doublets), compared to the conventional majority schemes. 
Thus we selected this scenario for both "high BW" and "low BW" approaches.

After the analysis steps described in  the previous section, the number of well-reconstructed events for both camera solutions is examined.
The number of events for high BW approach, normalized to the number of events for low BW system is shown on Fig.~\ref{SensitivityRatio_fig}. The ratio  curves for the nominal 
and the high (4.5 times brighter)  light of the night sky level  are depicted with circles and triangles accordingly.
As it can be seen, the high BW design provides slightly more events for further analysis up to the 1 TeV energies, gaining more prominently in the near-threshold energy region. 
For the LST subarray the gain of exploiting the high BW camera solution is substantial for all relevant energies (below 100 GeV).
Moreover, the gain for the high BW approach increases for the conditions of the bright night sky background and 
extends to the whole energy range .  


\section{Summary}

Lowering energy threshold and improving sensitivity
of large and medium CTA telescope size sections will be beneficial for the long-term monitoring of AGNs, the detectability of distant AGNs and gamma-ray bursts, pulsars and for the variability studies.
The R \& D work with novel photosensors and high bandwidth front-end camera electronics performed by the CTA consortium~\cite{bib:CTA, bib:ctamc} is very promising for improving the array performances. 
We extensively studied the impact of various trigger schemes and the rapidity of the front-end electronics on CTA 
gamma acceptances. The study on flux sensitivity and other key performances are ongoing. 

The observed lowering of energy threshold and improvement of gamma acceptances  for high bandwidth designs is moderate. However, there is a room for improvement on the
analysis side (e.g. using pixel-by-pixel image templates~\cite{bib:Model2D} and / or DISP-like~\cite{bib:DISP} methods etc.), which would further clarify the obtained performance difference between  two approaches. 
Next steps of our work
will incorporate some of these advanced image analysis methods, as well the next round of CTA simulations, which account for updated parameters of CTA candidate photosensors,
front-end electronics and other detector components.


\vspace*{0.5cm}
\footnotesize{{\bf Acknowledgment:}{We gratefully acknowledge financial
support from the agencies and organisations listed in this
page: http://www.cta-observatory.org/?q=node/22.}}

\end{document}